\documentclass[10pt,conference]{IEEEtran}

\usepackage{cite}
\usepackage{amsmath,amssymb,amsfonts}
\usepackage{algorithm}
\usepackage[noend]{algpseudocode}

\usepackage{caption}
\usepackage{subcaption}
\usepackage{graphicx}
\usepackage{textcomp}
\usepackage{xcolor}
\usepackage{listings}
\usepackage{enumitem}
\setlist[enumerate]{leftmargin=1.2em)}
\setlist[itemize]{leftmargin=1.2em}
\usepackage{color, colortbl}
\usepackage[most]{tcolorbox}
\usepackage{tikz}
\makeatletter
\algnewcommand{\LineComment}[1]{\Statex \hskip\ALG@thistlm \(\triangleright\) #1}
\makeatother
\algrenewcommand\algorithmicindent{1em}

\def\BibTeX{{\rm B\kern-.05em{\sc i\kern-.025em b}\kern-.08em
    T\kern-.1667em\lower.7ex\hbox{E}\kern-.125emX}}

\newtcolorbox{promptMSG}[3][]{%
    enhanced,
    attach boxed title to top right={yshift=-\tcboxedtitleheight},
    boxed title style={
        size=small,
        colback=gray!20,
        colframe=gray!20,
        sharp corners=downhill,
        arc=.5cm,
        top=1mm,bottom=1mm,left=1mm,right=1mm},
    fonttitle=\color{black}\itshape,
    colframe=#3,
    colback=#3,
    top=\tcboxedtitleheight,
    bottom=0mm,
    sharp corners=downhill,
    arc=.5cm,
    title={#2},#1
}

\lstdefinelanguage{JavaScript}{
   keywords={typeof, new, true, false, catch, function, return, null, catch, switch, var, if, in, while, do, else, case, break},
   keywordstyle=\color{blue}\bfseries,
   ndkeywords={class, export, boolean, throw, implements, import, this},
   ndkeywordstyle=\color{darkgray}\bfseries,
   identifierstyle=\color{black},
   sensitive=false,
   comment=[l]{//},
   morecomment=[s]{/*}{*/},
   commentstyle=\color{purple}\ttfamily,
   stringstyle=\color{red}\ttfamily,
   morestring=[b]',
   morestring=[b]"
}
 
\newcommand{\code}[1]{\texttt{\small#1}}
\newcommand{\name}{Treefix}

\begin{document}

% \title{\name{}: \underline{LL}M-Based \underline{L}earning-Guided Execution}
\title{\name{}: Enabling Execution with a \underline{Tree} of Pre\underline{fix}es}

\author{\IEEEauthorblockN{Beatriz Souza}
\IEEEauthorblockA{\textit{University of Stuttgart} \\
Stuttgart, Germany \\
beatrizbzsouza@gmail.com}
\and
\IEEEauthorblockN{Michael Pradel}
\IEEEauthorblockA{\textit{University of Stuttgart} \\
Stuttgart, Germany \\
michael@binaervarianz.de}
}

% \author{Beatriz Souza}
% \affiliation{%
%   \institution{University of Stuttgart}
%   \city{Stuttgart}
%   \country{Germany}}
% \email{beatrizbzsouza@gmail.com}

% \author{Michael Pradel}
% %\orcid{0000-0003-1623-498X}
% \affiliation{%
%   \institution{University of Stuttgart}
%   \city{Stuttgart}
%   \country{Germany}}
% \email{michael@binaervarianz.de}

\maketitle

\begin{abstract}
The ability to execute code is a prerequisite for various dynamic program analyses.
Learning-guided execution has been proposed as an approach to enable the execution of arbitrary code snippets by letting a neural model predict likely values for any missing variables.
Although state-of-the-art learning-guided execution approaches, such as LExecutor, 
can enable the execution of a relative high amount of code, they are limited to 
predicting a restricted set of possible values and do not use any feedback from previous executions to execute even more code.
This paper presents \name{}, a novel learning-guided execution approach that leverages LLMs to iteratively create code prefixes that enable the execution of a given code snippet.
The approach addresses the problem in a multi-step fashion, where each step uses feedback about the code snippet and its execution to instruct an LLM to improve a previously generated prefix.
This process iteratively creates a tree of prefixes, a subset of which is returned to the user as prefixes that maximize the number of executed lines in the code snippet.
In our experiments with two datasets of Python code snippets, \name{} achieves 25\% and 7\% more coverage relative to the current state of the art in learning-guided execution, covering a total of 84\% and 82\% of all lines in the code snippets.
\end{abstract}

\section{Introduction}

Executing code is essential for reasoning about the runtime behavior of code, e.g., in a dynamic program analysis, when extracting runtime data to train a model, or when trying to understand code during manual debugging. 
However, getting code to actually run is challenging, both at the small and the large scale.
At the small scale, individual code snippets, e.g., found in documentation or online forums, may contain undefined variables, functions, or classes, which prevent the code from executing. 
At the large scale, setting up a complex project is often difficult due to the diversity of build systems, missing dependencies, configuration issues, etc.
Even when a project is perfectly set up, executing a specific code location will require a specific set of inputs, which may not be available.

To enable the execution of arbitrary code snippets, either stand-alone snippets that simply are not executable on their own or code snippets extracted from a larger project, researchers have proposed \emph{learning-guided execution}~\cite{lexecutor}.
The basic idea to predict likely values for any missing variables in a code snippet with a neural model, preventing the execution from getting stuck.
Learning-guided execution has numerous applications because it enables dynamic analysis of code snippets in isolation.
Since its introduction in 2023, the community has started to explore several applications, such as detecting runtime type errors~\cite{Xue2024}, reproducing bugs~\cite{Hayet2024}, and checking whether a code change preserves the semantics~\cite{ChangeGuard2024}.
Learning-guided execution may also serve as a mechanism for validating code or code changes produced by an auto-programming technique, e.g., LLM-based code completion~\cite{Chen2021,Liu2023a} and code editing~\cite{Wei2023,Gupta2023,Dilhara2024}.
The ability of learning-guided execution to execute arbitrary code without requiring the full setup of the project should make it relatively easily applicable to a wide range of codebases. 

While the current state of the art in learning-guided execution, LExecutor~\cite{lexecutor}, can enable the execution of a relative high amount of code, it suffers from two key limitations:
(1) LExecutor predicts values sampled from a \textit{limited set of runtime values}.
Specifically, their neural model predicts one out of 23 abstract values, such as ``non-empty string'', which then are concretized into a hard-coded concrete value, such as \code{"a"}.
These values may not match realistic values, as they would occur in a real execution of the given code snippet, and they may not be diverse enough to reach all branches.
(2) LExecutor is designed for and evaluated on the task of executing a code snippet \emph{once}, which may not be sufficient to cover all branches.
Furthermore, executing the snippet only once prevents the approach from leveraging feedback from previous executions to improve the environment in which the given code snippets gets executed.
Overall, these two limitations curtail the effectiveness of the state-of-the-art approach at covering the lines in a given code snippet.

This paper presents \name{}, a novel learning-guided execution approach that introduces several ideas.
First, instead of training a custom model to predict abstract values, \name{} builds upon a large language model (LLM) to predict code that constructs concrete values.
We refer to the code that constructs these values as \emph{prefixes}, because the code gets prepended to the given code snippet before executing them together.
By generating code prefixes, \name{} can create a much larger and more diverse set of values, including domain-specific strings, complex objects, and even values returned from third-party libraries.
Second, \name{} reasons about the problem in a multi-step fashion, where each step uses feedback about the code snippet and its execution to instruct the LLM to improve a previously generated prefix.
The prefixes generated by \name{} form a tree, where each node represents a prefix and an edge represents a refinement of a prefix in the next step.
Finally, \name{} produces not only a single execution, but also returns a minimal set of prefixes that maximize the number of cumulatively executed lines in the code snippet.

The approach consists of three fully automated steps, which are designed to mimic the way a human may approach the problem of enabling the execution of a code snippet.
\begin{itemize}
    \item Step 1: Statically identify undefined references in the given code snippet and then query the LLM for prefixes that initialize them.
    \item Step 2: Execute the code snippet with the prefixes, observe any runtime errors that may occur, and then query the LLM for refined prefixes that address these errors.
    \item Step 3: Execute the code snippet with the refined prefixes, keep track of lines that are not yet covered, and then query the LLM for prefixes aimed at covering these lines. 
\end{itemize}

We evaluate \name{} by applying it to two sets of code snippets from prior work~\cite{lexecutor}: functions extracted from popular open-source projects and
code snippets extracted from Stack Overflow posts. As baselines, the evaluation compares \name{} with six alternative
approaches for executing arbitrary code snippets: regular execution, a 
state-of-the-art type predictor~\cite{type4Py}, a state-of-the-art function-level test generator~\cite{pynguin}, LExecutor~\cite{lexecutor}, Incompleter~\cite{Hayet2024}, and SelfPiCo~\cite{Xue2024}. 
We show that \name{} enables the execution of 84\% and 82\% of all lines in the open-source code and Stack Overflow snippets, respectively, 
which improves over the best baseline by 25\% and 7\%. Moreover, we show that each step of our approach contributes to its effectiveness and that \name{} produces,
on the evaluated snippets, 16528 unique values, which is 16505 more values than LExecutor. In case studies we find that \name{}'s ability to generate 
adequate imports, complex objects, and diverse primitive values contribute the most to \name{}'s improvements over LExecutor.

In summary, our contributions are as follows:

\begin{itemize}
    \item We propose \name{}, a multi-step learning-guided execution approach leveraging LLMs for enabling code execution.
    \item Through experimentation on two datasets, we show that \name{} can substantially improve code coverage 
    and is superior to the state-of-the-art approaches.
    % \item We conduct experiments to thoroughly evaluate \name{} and its design decisions and report insights from them.
    \item We release our open-source implementation of \name{} upon acceptance of the paper.
\end{itemize}

\begin{figure}
    \centering
    \begin{subfigure}[b]{\linewidth}
        \centering
        \includegraphics[width=0.95\textwidth]{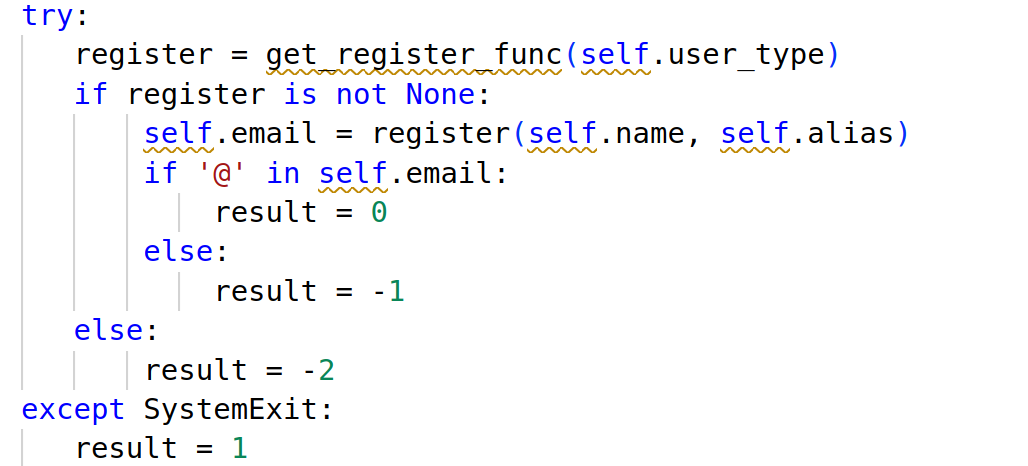}
        % \vspace{-1.7em}
        \caption{Python code to execute.}
        \label{fig:running_example_snippet}
    \end{subfigure}
    
    \vspace{1em}
    \begin{subfigure}[b]{\linewidth}
        \centering
        \includegraphics[width=0.95\textwidth]{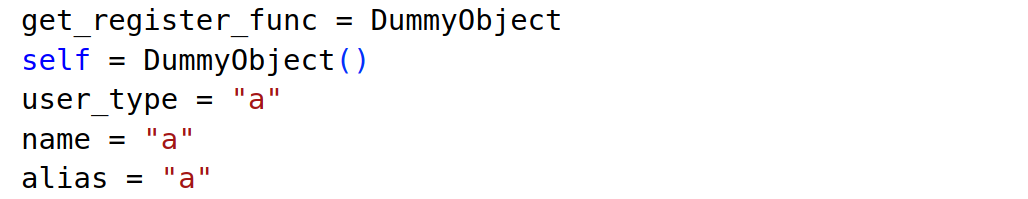}
        % \vspace{-1.7em}
        \caption{Values predicted by LExecutor~\cite{lexecutor}.}
        \label{fig:running_example_lexecutor_prediction}
    \end{subfigure}

    \vspace{1em}
    \begin{subfigure}[b]{\linewidth}
        \centering
        \includegraphics[width=0.95\textwidth]{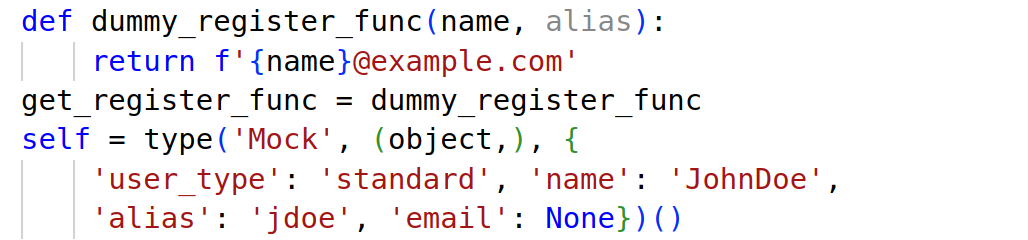}
        % \vspace{-1.7em}
        \caption{Prefix from step~1 of \name{} (undefinedness guidance).}
        \label{fig:running_example_l3_initial_prediction}
    \end{subfigure}

    \vspace{1em}
    \begin{subfigure}[b]{\linewidth}
        \centering
        \includegraphics[width=0.95\textwidth]{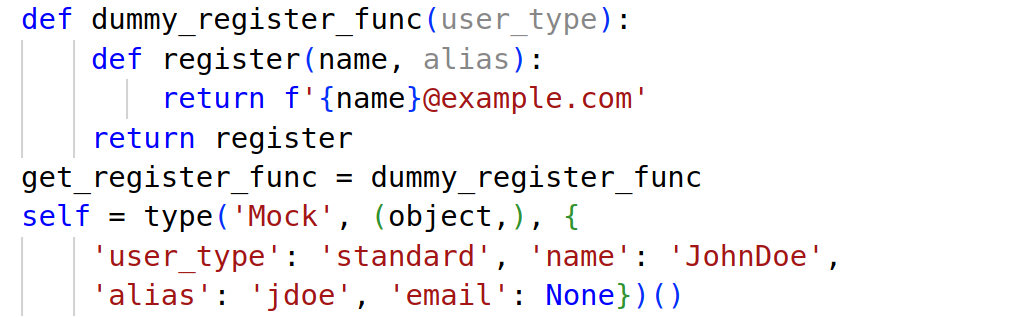}
        % \vspace{-1.7em}
        \caption{Prefix from step~2 of \name{} (error guidance).}
        \label{fig:running_example_l3_refine_prediction}
    \end{subfigure}

    \vspace{1em}
    \begin{subfigure}[b]{\linewidth}
        \centering
        \includegraphics[width=0.95\textwidth]{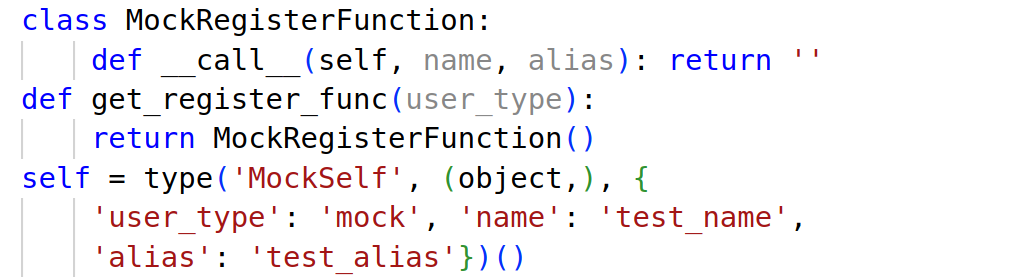}
        % \vspace{-1.7em}
        \caption{One prefix of step~3 of \name{} (coverage guidance).}
        \label{fig:running_example_treefix_guide_prediction_1}
    \end{subfigure}

    \vspace{1em}
    \begin{subfigure}[b]{\linewidth}
        \centering
        \includegraphics[width=0.95\textwidth]{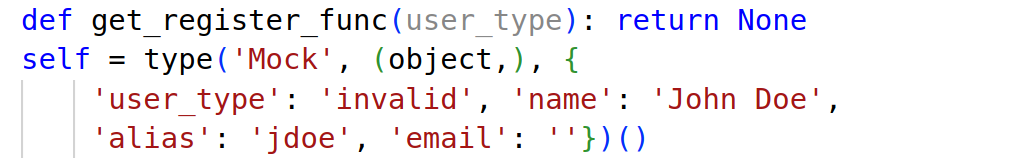}
        % \vspace{-1.7em}
        \caption{Another prefix of Step~3 of \name{} (coverage guidance).}
        \label{fig:running_example_treefix_guide_prediction_2}
    \end{subfigure}

    \vspace{1em}
    \begin{subfigure}[b]{\linewidth}
        \centering
        \includegraphics[width=0.95\textwidth]{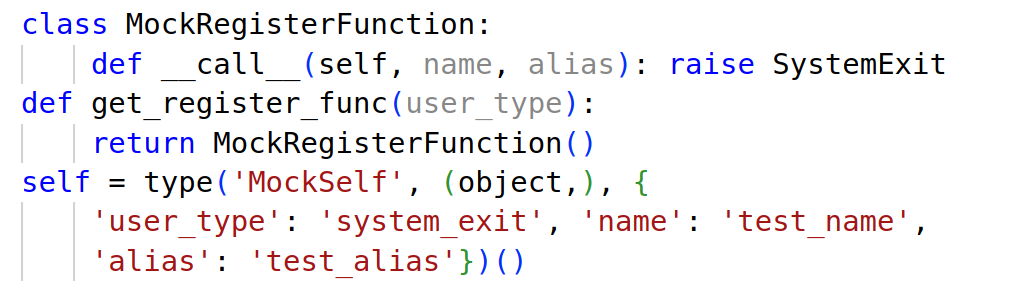}
        % \vspace{-1.7em}
        \caption{Yet another prefix of Step~3 of \name{} (coverage guidance).}
        \label{fig:running_example_treefix_guide_prediction_3}
    \end{subfigure}

        \caption{Example of code to execute and predicted values.}
        \label{fig:runtime_values}
        \vspace{-1.7em}
\end{figure}

\section{Motivating Example}

To motivate our work and illustrate the challenges of enabling code execution, consider the example in Figure~\ref{fig:running_example_snippet}.
The code could be part of a complex project or a code snippet extracted from an online resource.
Our goal is to execute this code snippet while covering as many of its lines as possible.
However, executing this code snippet is challenging due to various missing pieces of contextual information.
At first, the code tries to read the attribute \code{self.user\_type}, but \code{self} does not exist, which will cause the code to crash.
In case \code{self} and \code{self.user\_type} were defined, the code would try to call the function \code{get\_register\_func}, which is also undefined, 
as are the attributes \code{self.name} and \code{self.alias}.
Beyond undefined variables and functions, covering all code lines in this snippet is even more challenging due to the multiple branches and the \code{try} and \code{except} blocks.
To cover all lines, a single execution is not enough: A non-exceptional execution would miss out on the code in the \code{except} block, while an exceptional execution would miss out on any lines after the line that raises an exception in the \code{try} block.

The current state-of-the-art approach for executing arbitrary Python code snippets is LExecutor~\cite{lexecutor}, which has introduced the concept of learning-guided execution.
To enable the execution of arbitrary code snippets, LExecutor first trains a neural model that learns from real executions what values are typically used in a given code context.
LExecutor then modifies the code snippet so that whenever it would usually refer to a non-existing value and crash, the code instead queries the model to predict the most likely kind of value.
The approach then concretizes this kind of value, injects it into the running code, and continues the execution.
For example, a predicted kind of value could be ``non-empty string'', which LExecutor concretizes into a string \code{"a"}.

While LExecutor enables the execution of a relative high amount of code, it suffers from two keys limitations:
\begin{itemize}
    \item \emph{Restricted set of runtime values}.
    The kinds of values that LExecutor predicts are restricted to a fixed set of 23 abstractions, such as ``non-empty string'', ``negative integer'', and ``empty list''.
    For each of them, LExecutor has a single concrete value, such as \code{"a"}, \code{-1}, or \code{[]}, respectively.
    That is, the injected values may not match realistic values, as they would occur in a real execution of the given code snippet, and they may not be diverse enough to reach all branches.
    For the example in Figure~\ref{fig:running_example_snippet}, in case LExecutor predicts that the value assigned to \code{self.email} is a ``non-empty string'', the concrete value assigned to \code{self.email}
    would be \code{"a"}. In this case, it will not be able to execute the code in the second \code{if} statement, which checks if \code{"@"} is in \code{self.email}.
    Moreover, \code{"@"} would not be present in any other concrete value used by LExecutor. 
    \item \emph{Single execution}.
    LExecutor is designed for and evaluated on the task of executing a code snippet once, which may not be sufficient to cover all branches.
    In particular, any code snippet that contains mutually exclusive branches, such as two \code{if-else} branches without any loop or recursion around it, 
    such as in Figure~\ref{fig:running_example_snippet}, cannot be fully covered in a single execution.
\end{itemize}

Figure~\ref{fig:running_example_lexecutor_prediction} shows the values predicted by LExecutor for the code snippet in Figure~\ref{fig:running_example_snippet}.
Their approach predicts the return value of \code{get\_register\_func} to be a ``callable'', and hence, its concrete value is a \code{DummyObject} class,
which is assigned to the \code{register} variable.
As \code{register} is not \code{None}, the execution proceeds to the assignment of \code{self.email}.
LExecutor predicts \code{self} to be an ``object'', i.e., the concrete value is an instance of the \code{DummyObject} class.
Moreover, LExecutor predicts \code{self.user\_type}, \code{self.name}, and \code{self.alias} to be ``non-empty strings'', so they all get the concrete value \code{"a"}.
When \code{register} is called, a \code{DummyObject} instance is assigned to \code{self.email}.
This way, when the code tries to check if \code{"a"} is in \code{self.email}, a 
``TypeError: 'DummyObject' object is not iterable'' exception is raised, crashing the code execution.
As a result, only the first three lines 
in the \code{try} block are executed, but neither any of the remaining branches nor the \code{except} block, giving a line coverage of only 30\%.

\section{Approach}

The following presents \name{}, a multi-step, LLM-based approach to enable the execution of arbitrary code snippets.
We address the limitations of the state-of-the-art learning-guided execution approach, LExecutor, as follows.
Instead of predicting a restricted set of possible values, we leverage LLMs to predict code prefixes that can produce arbitrary values, such as domain-specific strings, complex objects, and even values returned from third-party libraries.
By prepending such a code prefix to the given code snippet, \name{} significantly increases the likelihood of executing the code snippet with realistic values that reach branches guarded by non-trivial conditions.
To address the limitation of using only a single execution, our approach creates a set of prefixes that iteratively increase the number of executed lines in the code snippet.
Finally, the approach yields a set of prefixes that complement each other in terms of coverage, e.g., by executing different branches of the code snippet.

\subsection{Problem Statement}

Before presenting \name{}, we start by defining the problem that we address.
The input to our approach is a syntactically valid piece of code, which we refer to as a \emph{code snippet} $s$.
The goal is to execute the code snippet, one or more times, to maximize the number of successfully executed lines.
To enable the execution of the code snippet, we generate one or more \emph{prefixes} $p$, where a prefix is a syntactically valid piece of code that consists of two parts: a possibly empty list of import statements and a possibly empty list of assignment statements that initialize variables used in $s$.

Because a single prefix $p$ may be insufficient to execute all lines in $s$, we aim to generate a set $P$ of prefixes that maximizes the number of executed lines in $s$.
We refer to those lines of $s$ that are executed when running $p+s$ as the \emph{coverage} achieved by $p$, and to those lines of $s$ that are executed when running $p+s$ for each $p \in P$ as the \emph{cumulative coverage}.
Using this terminology, the goal is to maximize the cumulative coverage of $s$ by generating the prefixes $P$, where one $p_{best} \in P$ will have the highest coverage achieved by any individual prefix.
The set $P$ and the single-best prefix $p_{best}$ are useful for different usage scenarios.
For example, $p_{best}$ can be used to understand the execution of $s$, e.g., by inspecting the execution in an interactive debugger.
Instead, $P$ can be used to dynamically analyze different executions of $s$, e.g., to detect behavioral anomalies.

While related, the problem addressed here differs from fuzzing~\cite{afl2013,Boehme2019,icse2024-Fuzz4All} and test case generation~\cite{pynguin, CodaMosa, coverUp}.
One difference is the assumptions about the given code to execute.
Both fuzzing and test case generation assume to have a complete and fully installed project, i.e., without any missing code and with all dependencies set up and ready to run.
In contrast, the code snippets that \name{} aims to execute are incomplete, e.g., due to undefined variables and functions, and may require additional dependencies to be installed.
Another difference is the interface used to provide values for the code to use.
Fuzzing typically provides values at the application interface, e.g., as command line arguments, and test case generation typically provides values at the function or method interface.
Instead, \name{} aims to fill in values that are missing at an arbitrary point in the given code snippet, without any well defined interface for providing these values.

\subsection{Overview and Running Example}

To address the problem of generating prefixes that maximize the cumulative coverage of a code snippet, \name{} reasons about the problem in a multi-step fashion.
The basic idea is to iteratively generate and refine prefixes until reaching full coverage or exceeding a configurable budget.
To generate and refine prefixes, \name{} queries an LLM with information obtained by statically and dynamically reasoning about the code snippet and already generated prefixes.
The approach provides three kinds of information to the model:
\emph{undefinedness guidance}, i.e., variables, attributes, and methods that are undefined in the code snippet;
\emph{error guidance}, i.e., runtime errors observed when executing the code snippet with a specific prefix; and
\emph{coverage guidance}, i.e., lines in the code snippet that are not yet covered by the currently known prefixes.

\begin{figure}[t]
    \includegraphics[width=\linewidth]{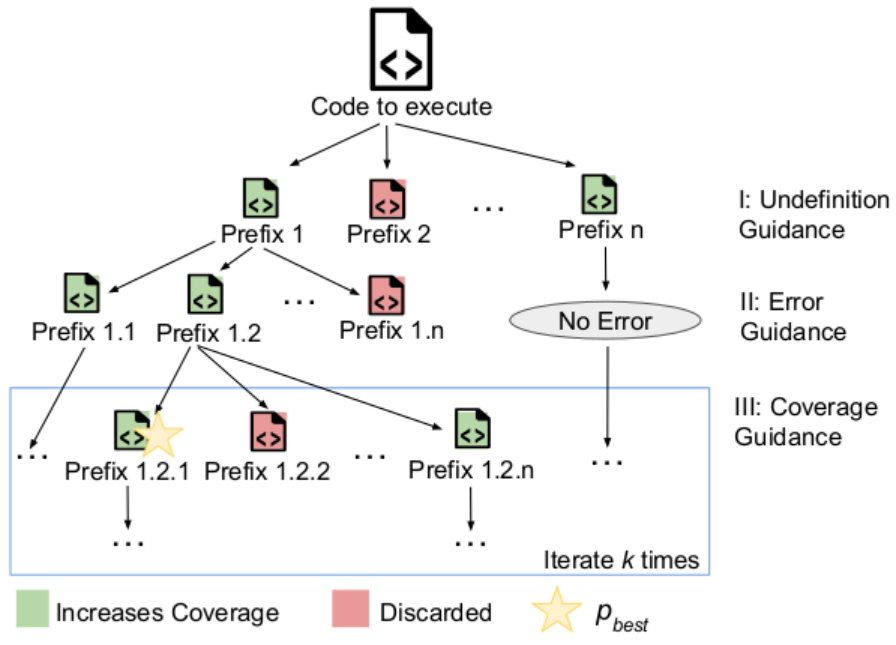}
    \caption{Tree of prefixes generated in different steps of \name{}. Prefixes highlighted in green are added to $P$. The best prefix $p_{best}$ is highlighted with a star.}
    \label{fig:tree}
\end{figure}

The prefixes generated by \name{} form a tree, as illustrated in Figure~\ref{fig:tree}.
The root is the given code snippet and each of the other nodes represents a prefix.
An edge represents a refinement of a prefix in the next step.
In addition to the root, the nodes in the tree are grouped into three levels, which correspond to the three kinds of guidance provided to the LLM.
The approach iteratively generates the tree of prefixes in a breadth-first manner.
That is, \name{} first generates prefixes based on undefinedness guidance, then refines them based on error guidance, and finally further refines the prefixes based on coverage guidance.
During this process, the approach keeps track of the prefixes $S$ that maximize cumulative coverage, highlighted in green in the figure, and of the single-best prefix $p_{best}$ that achieves the highest coverage, highlighted with a star.

The following illustrates the approach on the motivating example from Figure~\ref{fig:running_example_snippet}.

\paragraph*{Undefinedness guidance}
The first step is to statically identify any undefined values in the code and ask the LLM to predict code that initializes them.
Figure~\ref{fig:running_example_l3_initial_prediction} shows an example of  prediction made in the first step of \name{}.
The LLM predicts \code{get\_register\_func} to be the also predicted \code{dummy\_register\_func} function, which receives two arguments and returns 
a string.
Moreover, the model predicts \code{self} to be a \code{Mock} object containing the attributes accessed in the snippet.
While these values look legitimate at first sight, executing the prefix and the snippet shows that, when \code{get\_register\_func} is called on the first line 
of the \code{try} block, a ``TypeError: dummy\_register\_func() missing 1 required positional argument: 'alias''' exception is raised.

\paragraph*{Error guidance}
The second step of \name{} uses the feedback from the execution of the code snippet to refine the prefix, providing the invaluable information provided in the error message to the LLM.
Figure~\ref{fig:running_example_l3_refine_prediction} shows the refined prefix obtained in this step.
Notice that only the value of \code{dummy\_register\_func} changed. Now,
the LLM predict \code{dummy\_register\_func} to return another function,
which receives two arguments and returns a string containing \code{"@"}.
Given this prefix, the code snippet runs without raising any exceptions.
The execution covers the two \code{if} blocks, covering 60\% of the lines.

\paragraph*{Coverage guidance}
To further increase the coverage of the given code snippet, the third step of \name{} guides the LLM toward prefixes that execute those lines that were not executed in the previous steps.
For our example, those lines are the \code{except} and \code{else} blocks. 
Figures~\ref{fig:running_example_treefix_guide_prediction_1},~\ref{fig:running_example_treefix_guide_prediction_2}, and~\ref{fig:running_example_treefix_guide_prediction_3} 
show three prefixes obtained in the third step. 
The prefixes contain a value for \code{self.email} without \code{"@"}, a \code{register} set to \code{None}, and a statement that will raise a \code{SystemExit} exception, respectively.
The union of these prefixes successfully executes all lines in the code snippet, achieving 100\% line coverage.

\subsection{Main Algorithm}

\begin{algorithm}[t]
    \caption{Main algorithm of \name{}.}
    \label{alg:workflow}
    \begin{algorithmic}[1]
        \Require Code snippet $s$, number $n$ of prefixes to generate per prompt, number $k$ of coverage guidance attempts
        \Ensure Set $P$ of prefixes that maximize cumulative coverage and $p_{best} \in P$

        \State $P, p_{best}$ $\leftarrow \emptyset$, None

        \LineComment{\textbf{Step 1: Undefinedness guidance}}
        \State $V_\mathit{undefined}$ $\leftarrow$ \Call{getUndefinedRefs}{$s$}\label{l:step1_start}
        \State $\mathit{prompt_1}$ $\leftarrow$ \Call{genPrompt1}{$s$, $V_\mathit{undefined}$}
        \State $P_1$ $\leftarrow$ \Call{queryLLM}{$prompt_1$, $n$}
        \State $\mathit{execRes}$, $\mathit{covRes}$ $\leftarrow$ \Call{execute}{$P_1$, $s$}
        \State $P, p_{best} \leftarrow$ \Call{updatePrefixes}{$P$, $p_{best}$, $P_1$, $\mathit{covRes}$}\label{l:step1_end}

        \LineComment{\textbf{Step 2: Error guidance}}
        \State $P_\mathit{error} \leftarrow$ \Call{getErrorPrefixes}{$P_1$, $\mathit{execRes}$}\label{l:step2_start}

        \For {$p_\mathit{error}$ in $P_\mathit{error}$}
            \State $\mathit{prompt}_2$ $\leftarrow$ \Call{genPrompt2}{$s$, $p_\mathit{error}$}
            \State $P_2$ $\leftarrow$ \Call{queryLLM}{$\mathit{prompt}_2$, $n$}
            \State $\mathit{execRes}$, $\mathit{covRes}$ $\leftarrow$ \Call{execute}{$P_2$, $s$}
            \State $P, p_{best} \leftarrow$ \Call{updatePrefixes}{$P, p_{best}, P_2, \mathit{covRes}$}
        \EndFor\label{l:step2_end}

        \LineComment{\textbf{Step 3: Coverage guidance}}
        \State $P_3 \leftarrow P_2$\label{l:step3_start}
        \While {\Call{cumulativeCov}{$P$} $< 1$ and $k > 0$}
            \State $\mathit{execRes}$, $\mathit{covRes}$ $\leftarrow$ \Call{execute}{$P_3$, $s$}
            \State $s_\mathit{annot} \leftarrow$ \Call{annotateUncoveredLines}{$s$, $\mathit{covRes}$}
            \State $\mathit{prompt}_3$ $\leftarrow$ \Call{genPrompt3}{$s_\mathit{annot}$}
            \State $P_3$ $\leftarrow$ \Call{queryLLM}{$prompt_3$, $n$}
            \State $\mathit{execRes}$, $\mathit{covRes}$ $\leftarrow$ \Call{execute}{$P_3$, $s$}
            \State $P, p_{best} \leftarrow$ \Call{updatePrefixes}{$P, p_{best}, P_3, \mathit{covRes}$}
            \State $k$ $\leftarrow$ $k$ - 1
        \EndWhile\label{l:step3_end}

        \State \textbf{return} $P$, $p_{best}$
    \end{algorithmic}
\end{algorithm}

Algorithm~\ref{alg:workflow} summarizes the main steps of the approach.
In addition to the code snippet $s$, the algorithm takes two parameters as input: the number $n$ of prefixes to generate per prompt with the LLM and the number $k$ of coverage guidance attempts. 
The output of the algorithm is the set $P$ of prefixes that maximize cumulative coverage and the single-best prefix $p_{best}$ that achieves the highest coverage.
The remainder of this section describes the three steps of the approach in detail.

% \name{} starts by initializing the prefix $p$ and a minimal set of prefixes $P$ as in Algorithm~\ref{alg:workflow} (Line 1).
% \name{} contains three steps:
% (I) Undefinedness Guidance (Section~\ref{sec:step_1}), to predict missing values in a code snippet and enable its execution;
% (II) Error Guidance (Section~\ref{sec:step_2}), to fix predicted values in step 1 that are incomplete and lead to execution errors;
% and (III) Coverage Guidance (Section~\ref{sec:step_3}), to predict values that enable the execution of paths that were not previously executed 
% with the values from steps I and II. Each step starts with a \textit{getter} function
% that gets relevant information which is used for creating a prompt and querying a LLM. The responses
% from the LLM are executed to make sure that the predicted code can be executed and to measure the coverage achieved in $s$. 
% When the predictions made in the current and previous steps are able to execute all the lines in the code snippet, 
% the execution of \name{} ends. Otherwise, the next step is executed. In the end, \name{} returns $p_{best}$ and $P$ (Line 21).

\subsection{Step 1: Undefinedness Guidance}
\label{sec:step_1}

Given the code snippet $s$, the first step of \name{} aims to predict a prefix that initializes any undefined references in $s$.
These steps are summarized in lines~\ref{l:step1_start} to~\ref{l:step1_end} of Algorithm~\ref{alg:workflow}.
To identify all undefined references in the given code snippet, \name{} calls \textsc{getUndefinedRefs}, which is based on static analysis. 
\textsc{getUndefinedRefs} first parses the source code of the code snippet into an AST. 
Then, within the AST, it identifies the scopes of all variables in the code. 
Finally, \textsc{getUndefinedRefs} iterates over each scope and then over each variable access within that scope. 
If a variable is being accessed but not defined in the current scope or any enclosing scopes, it is considered undefined.
For example, in a code snippet \code{a = b + foo()} the variables \code{b} and \code{foo} are undefined.
Besides undefined variables, \textsc{getUndefinedRefs} also identifies undefined attributes and methods.
For example, in \code{y = d.year - p.init()}, the attribute \code{d.year} and the method \code{p.init()} are undefined, besides the undefined variables \code{d} and \code{p}.
For any undefined variable, any of its attributes or methods are considered to be undefined as well.
To determine the undefined attributes and methods, \textsc{getUndefinedRefs} identifies the locations of the nodes of the undefined variables and 
visits the attributes and methods being used in $s$. Whenever the base object of an undefined attribute or method matches with one of the 
undefined variables, the undefined attribute or method name with its base object, e.g., \code{d.year}, is added to a list, which is returned 
by \textsc{getUndefinedRefs}.

\begin{figure}[t]
    \centering
    \captionsetup{justification=centering} % Center the caption

    % msg 1
    \begin{promptMSG}{1}{green!10!white}
    \vspace{-1cm}
    \hspace{-0.2cm}
    Provide self-contained and concrete Python values to initialize the undefined variables in the code snippet.
    \end{promptMSG}

    \vspace{-0.2cm}

    % msg 2
    \begin{promptMSG}{2}{blue!10!white}
    \vspace{-1cm}
    \hspace{-2cm}
    \begin{lstlisting}[language=Python,linewidth=\linewidth,basicstyle=\ttfamily\scriptsize,escapechar={|},
        escapeinside={/*\#}{\#*/},keywordstyle=\color{blue!75!black},commentstyle=\color{green!40!black}]
# begin code snippet
(see Figure 1a)
# end code snippet
    \end{lstlisting}
    \vspace{-0.3cm}
    \end{promptMSG}

    \vspace{-0.2cm}

    % msg 3
    \begin{promptMSG}{3}{yellow!10!white}
    \vspace{-1cm}
    \hspace{-2cm}
    \begin{lstlisting}[language=Python,linewidth=\linewidth,basicstyle=\ttfamily\scriptsize,escapechar={|},
        escapeinside={/*\#}{\#*/},keywordstyle=\color{blue!75!black},commentstyle=\color{green!40!black}]
# begin undefined variables
self
get_register_func
# end undefined variables
    \end{lstlisting}
    \vspace{-0.3cm}
    \end{promptMSG}

    \vspace{-0.2cm}

    % msg 4
    \begin{promptMSG}{4}{pink!10!white}
    \vspace{-1cm}
    \hspace{-2cm}
    \begin{lstlisting}[language=Python,linewidth=\linewidth,basicstyle=\ttfamily\scriptsize,escapechar={|},
        escapeinside={/*\#}{\#*/},keywordstyle=\color{blue!75!black},commentstyle=\color{green!40!black}]
# begin undefined attributes and methods
self.user_type
self.name
self.alias
# end undefined attributes and methods
    \end{lstlisting}
    \vspace{-0.3cm}
    \end{promptMSG}

    \vspace{-0.2cm}

    % msg 5
    \begin{promptMSG}{5}{orange!10!white}
        \vspace{-10cm}
        \hspace{-0.2cm}
    Response specification (see Figure~\ref{fig:response_specification})
    \end{promptMSG}

    \vspace{-0.2cm}

    \caption{Prompt for undefinedness guidance.} 
    \label{fig:step1_prompt}
\end{figure}

\begin{figure}[t]
    \centering
    \captionsetup{justification=centering} % Center the caption

    \begin{promptMSG}{~}{orange!10!white}
    \vspace{-0.5cm}
    \hspace{-0.2cm}

    Respond strictly with JSON. The JSON should be compatible with the TypeScript type ``Response'':

    \begin{lstlisting}[language=JavaScript,linewidth=\linewidth,basicstyle=\ttfamily\scriptsize,escapechar={|},
        escapeinside={/*\#}{\#*/},keywordstyle=\color{blue!75!black},commentstyle=\color{green!40!black}]
```ts
interface Response {
  // Python import statements, one string per import
  imports: string[];

  // Python code to initialize undefined variables, one string per variable
  initialization: string[];
}
```
    \end{lstlisting}
    \vspace{-0.3cm}
    
    \end{promptMSG}

    \vspace{-0.3cm}

    \caption{Response specification.}
    \label{fig:response_specification}
\end{figure}

Based on the statically determined set of undefined references, \name{} generates a prompt aimed at predicting code to define those references (\textsc{genPrompt1}).
Figure~\ref{fig:step1_prompt} shows the prompt (slightly modified for readability), which has the following structure:
1) a request to provide the missing values;
2) the code snippet $s$ with comments indicating its beginning and end;
3) the list of undefined references in $s$;
4) the list of undefined attributes and methods in $s$;
5) a specification of the expected response, as described in Figure~\ref{fig:response_specification}.
Next, \name{} calls \textsc{queryLLM}, which queries the model with the prompt to obtain $n$ prefixes.
The rationale for generating multiple prefixes for the prompt is to increase the diversity of the prefixes generated, and hence, the chance to find prefixes that successfully cover the code in $s$.

Given the prefixes returned by the LLM, which are stored in set $P_1$ in Algorithm~\ref{alg:workflow}, \name{} executes them and updates the coverage information.
Because this part of the approach is used for all three of \name{}'s steps, we describe it in more detail in Section~\ref{sec:execute}.
In short, \textsc{execute} post-processes each prefix, automatically installs any third-party dependencies required to execute the prefix, prepends the prefixes to the code snippet $s$, and then measures the line coverage achieved by it.
Finally, \name{} updates the set of prefixes $P$ and the single-best prefix $p_{best}$ based on the coverage achieved by the prefixes in $P_1$.

\subsection{Step 2: Error Guidance}
\label{sec:step_2}

The values predicted in step 1 of \name{} may be incomplete or contain values that lead to execution errors, 
as in Figure~\ref{fig:running_example_l3_initial_prediction}.
To fix any errors, the second step of \name{} uses the error messages observed during the execution of the prefixes generated in step 1 as feedback to refine any erroneous prefixes (lines~\ref{l:step2_start} to~\ref{l:step2_end} in Algorithm~\ref{alg:workflow}).
The helper function \textsc{getErrorPrefixes} checks the execution results of the prefixes in $P_1$ and returns those prefixes that raised an exception.
Each exception contains the exception type, message, and line number where it happened.

For each prefix that raises an exception, \name{} formulates a prompt (\textsc{genPrompt2}) aimed at predicting a fixed version of the prefix.
Figure~\ref{fig:step2_prompt} shows the prompt structure, which contains the following: 
1) a description of the problem; 
2) the exception type, message, and line number where it happened;
3) a statement of the task;
4) a response specification indicating the response content and its format, as described in Figure~\ref{fig:response_specification}. 
Similar to step 1, \name{} queries the LLM with the prompt to obtain $n$ refined prefixes, which are stored in set $P_2$ in the algorithm.
Because LLMs tend to be more effective when given meaningful context, \name{} keeps the conversation history that has led to the erroneous prefix in step~1 as part of the prompt for step~2.
% In step 2, before each $prompt$, \name{} adds the prefix that caused the exception present in the $prompt$ to the conversation history and
% provides the conversation history when querying the model.
Finally, the approach executes the prefixes in $P_2$ and updates the coverage information (again, Section~\ref{sec:execute} will provide the details). 

\begin{figure}[t]
    % msg 1
    \begin{promptMSG}{1}{red!10!white}
    \vspace{-0.5cm}
    \hspace{-0.2cm}
    When trying to execute the code snippet with the provided imports and initialization, the following error happens:
    \end{promptMSG}

    \vspace{-0.2cm}

    % msg 2
    \begin{promptMSG}{2}{yellow!10!white}
    \vspace{-1cm}
    \hspace{-2cm}
    \begin{lstlisting}[language=Python,linewidth=\linewidth,basicstyle=\ttfamily\scriptsize,escapechar={|},
        escapeinside={/*\#}{\#*/},keywordstyle=\color{blue!75!black},commentstyle=\color{green!40!black}]
# begin error message
Execution error at line 14:
    register = get_register_func(self.user_type)
TypeError: dummy_register_func() missing 1 required positional argument: 'alias'
# end error message
    \end{lstlisting}
    \vspace{-0.3cm}
    \end{promptMSG}

    \vspace{-0.2cm}

    % msg 3
    \begin{promptMSG}{3}{green!10!white}
    \vspace{-0.6cm}
    %\hspace{-2cm}
    Provide a fixed version of the imports and initialization to solve the error and make the code snippet executable.
    \end{promptMSG}

    \vspace{-0.2cm}

    % msg 4
    \begin{promptMSG}{4}{orange!10!white}
    \vspace{-10cm}
    \hspace{-0.2cm}
    Response specification (see Figure~\ref{fig:response_specification})
    \end{promptMSG}

    \vspace{-0.2cm}

    \caption{Prompt for error guidance.}
    \label{fig:step2_prompt}
\end{figure}

\subsection{Step 3: Coverage Guidance}
\label{sec:step_3}

Steps 1 and 2 are often successfully at finding one or more prefixes that enable executing the given code snippet without errors.
However, there may still be lines in the code snippet that are not executed by any of the prefixes generated in the previous steps, 
such as the multiple branches in Figure~\ref{fig:running_example_snippet}.
To increase the coverage of the code snippet, the third step of \name{} aims to predict prefixes that exercise any not yet covered lines, as summarized in lines~\ref{l:step3_start} to~\ref{l:step3_end} in Algorithm~\ref{alg:workflow}.

The basic idea of step 3 is to iteratively generate additional prefixes until either reaching full coverage or exhausting a budget of $k$ attempts.
In each iteration, \name{} calls \textsc{annotateUncoveredLines}, which identifies all the lines in $s$ that have not been covered by the previous predictions and marks them using a special comment \lstinline{# uncovered}.
The approach then formulates a prompt aimed at predicting a prefix that exercises the uncovered lines.
Figure~\ref{fig:step3_prompt} shows the prompt structure, which contains:
1) a description of the problem; 
2) the annotated code;
3) a statement of the task;
4) the response specification. 
As in Steps 1 and 2, \name{} calls the $LLM$ to generate $n$ prefixes.
For each prefix, the approach executes it and updates the coverage information.

\begin{figure}[t]
    % msg 1
    \begin{promptMSG}{1}{red!10!white}
    \vspace{-0.5cm}
    \hspace{-0.2cm}
    When trying to execute the code snippet with the provided imports and initialization, the lines commented with ``uncovered'' are not executed.
    \end{promptMSG}

    \vspace{-0.2cm}

    % msg 2
    \begin{promptMSG}{2}{yellow!10!white}
    \vspace{-1cm}
    \hspace{-2cm}
    \begin{lstlisting}[language=Python,linewidth=\linewidth,basicstyle=\ttfamily\scriptsize,escapechar={|},
        escapeinside={/*\#}{\#*/},keywordstyle=\color{blue!75!black},commentstyle=\color{green!40!black}]
# begin code snippet
try:
    register = get_register_func(
        self.user_type)
    if register is not None:
        self.email = register(
            self.name, self.alias)
        if '@' in self.email:
            result = 0
        else:
            result = -1 # uncovered
    else:
        result = -2 # uncovered
 except SystemExit: # uncovered
    result = 1 # uncovered
# end code snippet
    \end{lstlisting}
    \vspace{-0.3cm}
    \end{promptMSG}

    \vspace{-0.2cm}

    % msg 3
    \begin{promptMSG}{3}{green!10!white}
    \vspace{-0.6cm}
    Provide a modified version of the imports and initialization to execute one of the uncovered paths in the code snippet.
    \end{promptMSG}

    \vspace{-0.2cm}

    % msg 4
    \begin{promptMSG}{4}{orange!10!white}
    \vspace{-10cm}
    \hspace{-0.2cm}
    Response specification (see Figure~\ref{fig:response_specification})
    \end{promptMSG}

    \vspace{-0.2cm}

    \caption{Prompt for coverage guidance.}
    \label{fig:step3_prompt}
\end{figure}

\subsection{Execution and Coverage Measurement}
\label{sec:execute}

A key component of \name{} is to obtain feedback by executing the prefixes generated by the LLM.
The following presents code execution and coverage measurement in more detail, which corresponds to the helper functions \textsc{execute} and \textsc{updatePrefixes} in Algorithm~\ref{alg:workflow}.
These helper functions are used in all three steps of \name{}.

\subsubsection{Installing Third-Party Dependencies}
\label{sec:install_dependencies}

The predicted prefixes may contain imports from dependencies that are not currently installed in the environment where \name{} is being executed.
Our approach automatically identifies and installs any missing dependencies.
To this end, we use \emph{pipreqs}, a Python library that identifies the dependencies based on the imports in the code.
For example, the predictions made by \name{} in Figure~\ref{fig:case_study_example_prediction} start with two \code{import} statements. 
In this case, \name{} identifies that \code{pandas} and \code{numpy} are dependencies and installs them.
Because installing dependencies is one of the most time-consuming processes of \name{}, we keep a shared environment between snippets.
This environment contains all installed dependencies and avoids repeatedly installing the same libraries.

\subsubsection{Post-Processing of Prefixes}
\label{sec:post_processing}

Some of the prefixes generated by the LLM may contain errors, yet other parts of the same prefix are useful.
\name{} post-processes the prefixes to heuristically remove any lines that raise an execution error.
Specifically, the approach iteratively attempts to execute each prefix up to 10 times and removes any lines that  raise an execution error.
If this process results in a prefix that runs without errors, it will be concatenated with the code snippet $s$ and executed to measure the coverage achieved, as described below.
Otherwise, the prefix is discarded.
Another problem is that the predicted prefixes may contain an infinity loop or take very long to run. 
To work around this problem, \name{} also removes prefixes that take more than 30 seconds to execute.

\subsubsection{Measure Coverage}
\label{sec:measure_coverage}

For all prefixes that, when executed on their own, neither raise an error nor time out, \name{} measures the coverage achieved when prepending the prefix to the code snippet.
To measure the coverage achieved by a prefix in $s$, \name{} uses the same strategy to measure coverage used by LExecutor.
It instruments $s$ by adding a call to a special function \code{\_l\_} after every 
line in the code. \code{\_l\_} receives a unique ID, which identifies the line above, as argument. Whenever \code{\_l\_} is called, 
the previous line was successfully executed. \name{} combines, in this order, the predicted and post-processed prefix with the instrumented version of 
$s$ into a program, and then executes it to record the executed line numbers. 
Unlike reports by popular coverage tools, this measurement considers a line ``covered'' only if the entire line executes without crashing.

\section{Evaluation}

We structure our evaluation along five research questions.
\begin{itemize}
    \item RQ1: How much code does an execution guided by \name{} cover, and how does it compare to prior work?
    \item RQ2: How do the three steps in \name{} contribute to its effectiveness?
    \item RQ3: Qualitatively, why does \name{} achieve different coverage results than existing work?
    \item RQ4: How diverse are the values predicted by \name{}?
    \item RQ5: What are the costs of executing code with \name{}?
\end{itemize}

\subsection{Experimental Setup}

\begin{table}[t]
    \centering
    \caption{Datasets used for evaluation.}
    \label{tab:eval_datasets}
    \begin{tabular}{lrr}
    \hline
    Dataset                 & \multicolumn{1}{r}{Snippets} & \multicolumn{1}{r}{LoC}  \\ \hline
    Open-source functions   & 1,000                         & 9,653                                                      \\
    Stack Overflow snippets & 462                          & 3,580                                                      \\ \hline
    Total                   & 1,462                        & 13,233                                                       \\ \hline
    \end{tabular}
\end{table}

\subsubsection{Benchmark Datasets}

We evaluate on two datasets, described in Table~\ref{tab:eval_datasets}, containing Python code snippets used in previous work~\cite{lexecutor}.
The \textit{Open-source functions} dataset contains 1,000 randomly selected functions from five large and diverse open-source Python projects.
The \textit{Stack Overflow snippets} dataset contains 462 code snippets from the answers to Python questions on Stack Overflow.

\subsubsection{Baselines}

We compare \name{} with six alternative approaches for executing arbitrary code snippets:
1) LExecutor~\cite{lexecutor}, for which we use the most effective variation, i.e., the fine-grained value abstraction, using top-1 predictions from the CodeT5 model.
2) SelfPiCo~\cite{Xue2024}, an LLM-based approach developed concurrently with this work, which guides code execution in an interactive loop.
We apply their approach to the datasets we use here, which is the same as in the LExecutor paper, but different from the dataset used in the SelfPiCo paper.
As their fine-tuned Code Llama model is not available, we use their approach with GPT-3.5, which they report to achieve similar performance than the version with Code Llama. 
3) Incompleter~\cite{Hayet2024}, a rule-based, feedback-driven approach. It measures coverage using \emph{coverage.py}, which -- unlike our coverage metric -- counts a line as covered even if that line crashes.
For a fair comparison, we modify Incompleter to measure coverage as we do for \name{} and all the other baselines, i.e., considering a line as covered if it is successfully executed.
We apply Incompleter to the datasets we use.
In addition to these three state-of-the-art baselines, we also consider the baselines that LExecutor has been compared with:
4) ``As Is'', i.e., trying to execute a code snippet as it is without making any value predictions.
5) Pynguin, a function-level test generator for Python~\cite{pynguin}. As in previous 
work~\cite{lexecutor}, for a fair comparison, we run Pynguin on a single function at a time, which contains only the code to execute.
6) Type4Py, a neural model that predicts types for all local variables, parameters, and return values~\cite{type4Py}. We concretize the predicted types as done for this baseline in previous work~\cite{lexecutor}.

\subsubsection{Evaluated Models}

We use OpenAI's latest flagship models: GPT-4o, the largest available model, and GPT-4o mini, a more lightweight and cheaper model.\footnote{https://platform.openai.com/docs/models} 

\subsubsection{Algorithm Parameters} 

We set the maximum number of completions to $n=10$ for each query to the model. 
Moreover, in step~3, we set the maximum number of iterations $k=10$. 
% Finally, whenever we check if a code snippet is executable, we limit the execution time to 30 seconds.

\subsection{RQ1: Effectiveness at Covering Code}

\begin{table*}[t]
    \caption{Effectiveness achieved by \name{} and baselines.}
    \label{tab:overall_effectiveness}
    \centering
    \resizebox{\textwidth}{!}{
    \begin{tabular}{lrrrrrrrlrr}
    \hline
    Approach       & \multicolumn{5}{c}{Coverage}                                                                                                                                             & \multicolumn{1}{l}{}     & \multicolumn{4}{c}{Full Execution Rate}                                                                                   \\ \cline{2-6} \cline{8-11} 
                   & \multicolumn{2}{c}{Open-source functions}                             & \multicolumn{1}{l}{}     & \multicolumn{2}{c}{Stack Overflow snippets}                           & \multicolumn{1}{l}{}     & \multicolumn{2}{c}{Open-source functions}        & \multicolumn{1}{l}{}     & \multicolumn{1}{c}{Stack Overflow snippets} \\ \cline{2-3} \cline{5-6} \cline{8-9} \cline{11-11} 
    Treefix (GPT4o)       & \cellcolor[HTML]{CCD7E6}$P$ = \textbf{0.84} & \cellcolor[HTML]{CCD7E6}$p_{best}$ = \textbf{0.76} & \cellcolor[HTML]{CCD7E6} & \cellcolor[HTML]{CCD7E6}$P$ = \textbf{0.82} & \cellcolor[HTML]{CCD7E6}$p_{best}$ = 0.72 & \cellcolor[HTML]{CCD7E6} & \multicolumn{2}{r}{\cellcolor[HTML]{CCD7E6}\textbf{0.69}} & \cellcolor[HTML]{CCD7E6} & \cellcolor[HTML]{CCD7E6}\textbf{0.71}                \\
    Treefix (GPT4o-mini) & \cellcolor[HTML]{E2E9F3}$P$ = 0.79 & \cellcolor[HTML]{E2E9F3}$p_{best}$ = 0.73 & \cellcolor[HTML]{E2E9F3} & \cellcolor[HTML]{E2E9F3}$P$ = 0.79 & \cellcolor[HTML]{E2E9F3}$p_{best}$ = \textbf{0.78} & \cellcolor[HTML]{E2E9F3} & \multicolumn{2}{r}{\cellcolor[HTML]{E2E9F3}0.62} & \cellcolor[HTML]{E2E9F3} & \cellcolor[HTML]{E2E9F3}0.66                \\
    SelfPiCo      & \multicolumn{2}{r}{\cellcolor[HTML]{EDEFF3}0.59}                      & \cellcolor[HTML]{EDEFF3} & \multicolumn{2}{r}{\cellcolor[HTML]{EDEFF3}0.75}                      & \cellcolor[HTML]{EDEFF3} & \multicolumn{2}{r}{\cellcolor[HTML]{EDEFF3}0.40} & \cellcolor[HTML]{EDEFF3} & \cellcolor[HTML]{EDEFF3}0.60                \\
    Incompleter      & \multicolumn{2}{r}{\cellcolor[HTML]{EDEFF3}0.51}                      & \cellcolor[HTML]{EDEFF3} & \multicolumn{2}{r}{\cellcolor[HTML]{EDEFF3}0.69}                      & \cellcolor[HTML]{EDEFF3} & \multicolumn{2}{r}{\cellcolor[HTML]{EDEFF3}0.35} & \cellcolor[HTML]{EDEFF3} & \cellcolor[HTML]{EDEFF3}0.53                \\
    LExecutor      & \multicolumn{2}{r}{\cellcolor[HTML]{EDEFF3}0.51}                      & \cellcolor[HTML]{EDEFF3} & \multicolumn{2}{r}{\cellcolor[HTML]{EDEFF3}0.65}                      & \cellcolor[HTML]{EDEFF3} & \multicolumn{2}{r}{\cellcolor[HTML]{EDEFF3}0.35} & \cellcolor[HTML]{EDEFF3} & \cellcolor[HTML]{EDEFF3}0.49                \\
    Type4Py        & \multicolumn{2}{r}{0.13}                                              &                          & \multicolumn{2}{r}{0.46}                                              &                          & \multicolumn{2}{r}{0.08}                         &                          & 0.32                                        \\
    Pynguin tests  & \multicolumn{2}{r}{0.04}                                              &                          & \multicolumn{2}{r}{-}                                                 &                          & \multicolumn{2}{r}{0.02}                         &                          & -                                           \\
    As Is          & \multicolumn{2}{r}{0.04}                                              &                          & \multicolumn{2}{r}{0.43}                                              &                          & \multicolumn{2}{r}{0.02}                         &                          & 0.30                                        \\ \hline
    \end{tabular}
    }
\end{table*}

Table~\ref{tab:overall_effectiveness} (left side) shows the line coverage achieved by \name{} and by the baselines on the two datasets.
We use the Wilcoxon signed-rank test to compare the significance of coverage differences between techniques, at p = 0.05.
For the open-source functions, on average, executing the code as it is and the type (Type4Py) predictor
cover only 4.1\% and 13.3\% of the lines, respectively. LExecutor and Incompleter increase the mean coverage to 51.6\%.
Then, SelfPiCo further increases the coverage to 59\%.
Finally, \name{} covers 76\% of the lines with $p_{best}$ and even 84\% with $P$, which is higher than all considered baselines (statistically significant) and an improvement of 25\% over SelfPiCo, i.e., the currently best available approach.
Comparing the two LLMs shows that using the larger GPT4o model is more effective than using GPT4o-mini.
Nevertheless, even with GPT4o-mini, \name{} is still significantly more effective than SelfPiCo.
On the Stack Overflow code snippets datasets, \name{} covers 82\% of the lines with $P$, an improvement of 7\% over SelfPiCo.

We also measure the full execution rate, i.e., how many of all code snippets achieve 100\% line coverage. 
The results are presented in Table~\ref{tab:overall_effectiveness} (right side).
Overall, \name{} outperforms the baselines and fully executes the code of 69\% of the open-source functions and of 71\% of the Stack Overflow code snippets.
In contrast, SelfPiCo fully executes the code of 62\% of all open-source functions and of 66\% of all Stack Overflow code snippets.

\subsection{RQ2: Design Choices}

\begin{table*}[]
    \caption{Effectiveness achieved on each step of \name{}.}
    \label{tab:effectiveness_per_step}
    \centering
    \begin{tabular}{llrrrrrrr}
    \hline
    Model                                                     & Dataset                         & \multicolumn{3}{c}{Coverage}                                                                        & \multicolumn{1}{c}{} & \multicolumn{3}{c}{Full Execution Rate}                                                             \\ \cline{3-5} \cline{7-9} 
                                                              &                                 & \multicolumn{1}{c}{I}        & \multicolumn{1}{c}{II}       & \multicolumn{1}{c}{III}               & \multicolumn{1}{c}{} & \multicolumn{1}{c}{I}        & \multicolumn{1}{c}{II}       & \multicolumn{1}{c}{III}               \\ \cline{1-5} \cline{7-9} 
    GPT4o                                                     & Open-source functions           & \cellcolor[HTML]{EDEFF3}0.72 & \cellcolor[HTML]{E2E9F3}0.78 & \cellcolor[HTML]{CCD7E6}\textbf{0.84} &                      & \cellcolor[HTML]{EDEFF3}0.54 & \cellcolor[HTML]{E2E9F3}0.61 & \cellcolor[HTML]{CCD7E6}\textbf{0.69} \\
                                                              & Stack Overflow snippets         & \cellcolor[HTML]{EDEFF3}0.73 & \cellcolor[HTML]{E2E9F3}0.77 & \cellcolor[HTML]{CCD7E6}\textbf{0.82} &                      & \cellcolor[HTML]{EDEFF3}0.59 & \cellcolor[HTML]{E2E9F3}0.64 & \cellcolor[HTML]{CCD7E6}\textbf{0.71} \\ \hline
    \begin{tabular}[c]{@{}l@{}}GPT4o\\ (mini)\end{tabular}    & Open-source functions           & \cellcolor[HTML]{EDEFF3}0.64 & \cellcolor[HTML]{E2E9F3}0.74 & \cellcolor[HTML]{CCD7E6}0.79          &                      & \cellcolor[HTML]{EDEFF3}0.46 & \cellcolor[HTML]{E2E9F3}0.56 & \cellcolor[HTML]{CCD7E6}0.62          \\
                                                              & Open-source functions  w/o pipreqs   & \cellcolor[HTML]{EDEFF3}0.60 & \cellcolor[HTML]{E2E9F3}0.70 & \cellcolor[HTML]{CCD7E6}0.74          &                      & \cellcolor[HTML]{EDEFF3}0.44 & \cellcolor[HTML]{E2E9F3}0.53 & \cellcolor[HTML]{CCD7E6}0.58          \\
                                                              & Stack Overflow snippets         & \cellcolor[HTML]{EDEFF3}0.70 & \cellcolor[HTML]{E2E9F3}0.72 & \cellcolor[HTML]{CCD7E6}0.79          &                      & \cellcolor[HTML]{EDEFF3}0.56 & \cellcolor[HTML]{E2E9F3}0.59 & \cellcolor[HTML]{CCD7E6}0.66          \\
                                                              & Stack Overflow snippets  w/o pipreqs & \cellcolor[HTML]{EDEFF3}0.67 & \cellcolor[HTML]{E2E9F3}0.71 & \cellcolor[HTML]{CCD7E6}0.77          &                      & \cellcolor[HTML]{EDEFF3}0.54 & \cellcolor[HTML]{E2E9F3}0.58 & \cellcolor[HTML]{CCD7E6}0.64          \\ \hline
    \end{tabular}
\end{table*}

\begin{figure}
    \centering
    \begin{subfigure}[b]{\linewidth}
        \centering
        \includegraphics[width=0.65\textwidth]{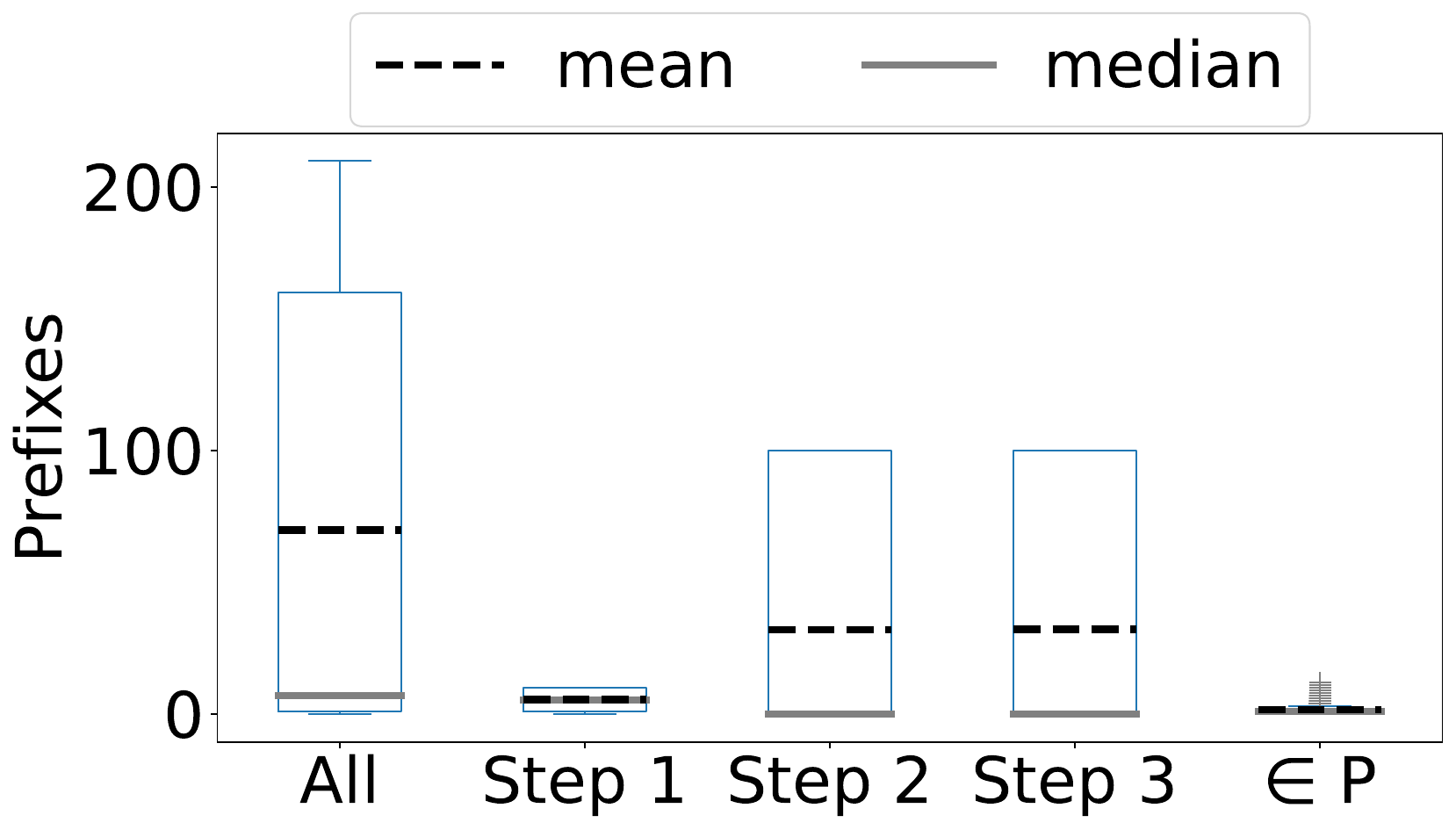}
        %\vspace{-1.7em}
        \caption{Open-source functions.}
        \label{fig:prefixes_exploration_a}
    \end{subfigure}

    \vspace{1em}
    \begin{subfigure}[b]{\linewidth}
        \centering
        \includegraphics[width=0.65\textwidth]{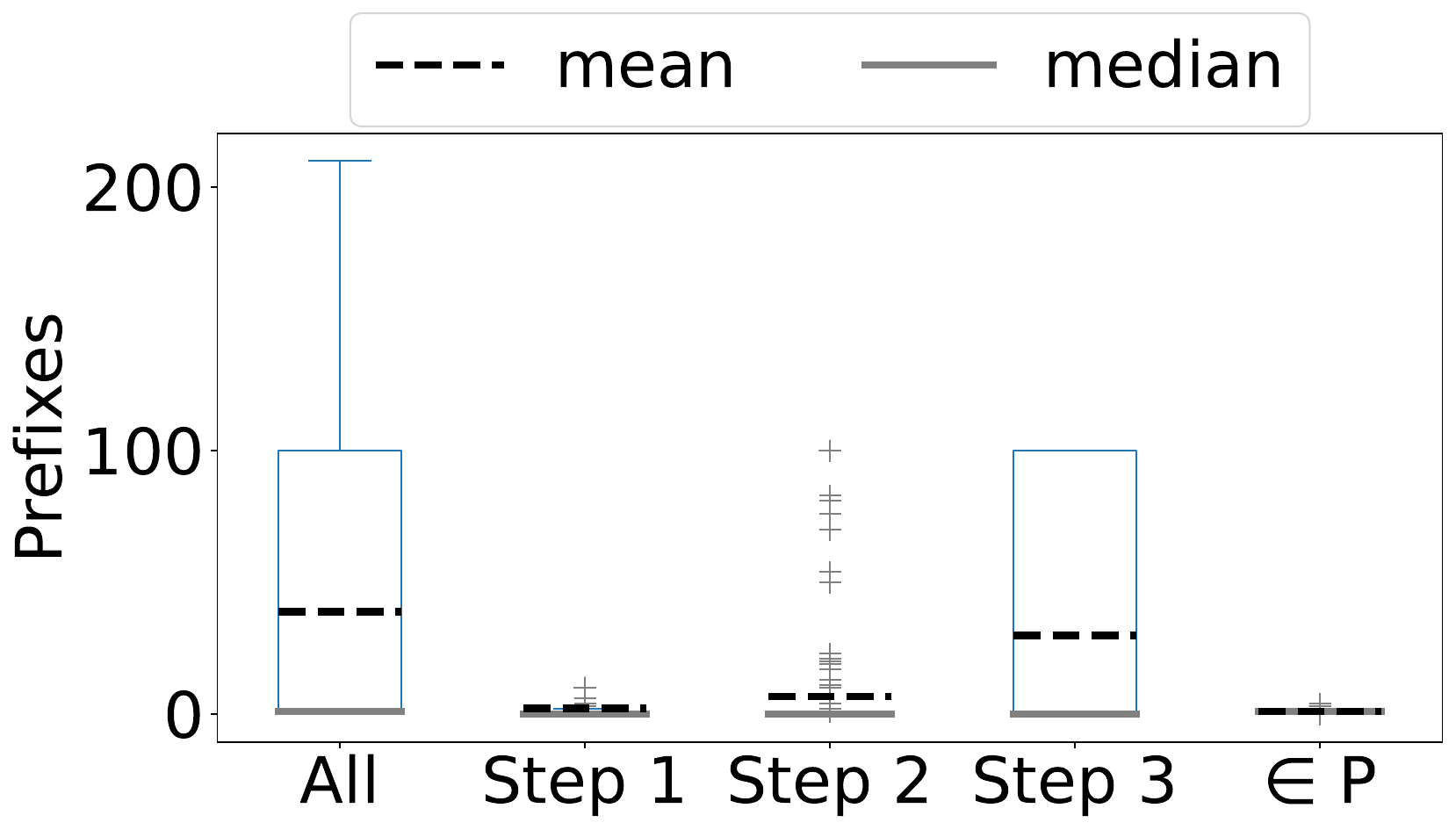}
        %\vspace{-1.7em}
        \caption{Stack Overflow snippets.}
        \label{fig:prefixes_exploration_b}
    \end{subfigure}
    \caption{Prefixes explored by \name{}.}
    \label{fig:prefixes_exploration}
    \vspace{-1.7em}
\end{figure}

\subsubsection{Effectiveness per Step}
\name{} uses three steps to enable code execution. We evaluate how each of these steps contributes to the improvements in effectiveness.
To this end, we measure the line coverage and full execution rate achieved after each step, again for both datasets and models.
Table~\ref{tab:effectiveness_per_step} presents the results. 
On average, \name{}, with either of the considered models, in the first step already achieves comparable or higher effectiveness than the best baseline, 
SelfPiCo (Table~\ref{tab:overall_effectiveness}).
Steps~2 and~3 consistently further increase effectiveness on both considered datasets,
regardless of the model.

\begin{table}[t]
    \caption{Snippets where $p_{best}$ is found in a specific step.}
    \label{tab:p_best_location}
    \centering
    \resizebox{0.7\columnwidth}{!}{
    \begin{tabular}{lcrr}
    \hline
    Dataset                 & \multicolumn{3}{c}{Step}                                                                                   \\ \cline{2-4} 
                            & I                                               & \multicolumn{1}{c}{II}      & \multicolumn{1}{c}{III}    \\ \hline
    Open-source functions   & \multicolumn{1}{r}{\cellcolor[HTML]{CCD7E6}\textbf{719}} & \cellcolor[HTML]{E2E9F3}129 & \cellcolor[HTML]{EDEFF3}70 \\
    Stack Overflow snippets & \multicolumn{1}{r}{\cellcolor[HTML]{CCD7E6}\textbf{356}} & \cellcolor[HTML]{EDEFF3}28  & \cellcolor[HTML]{E2E9F3}45 \\ \hline
    \end{tabular}
    }
\end{table}

To understand when \name{} typically finds the single-best prefix, Table~\ref{tab:p_best_location} shows the number of snippets where $p_{best}$ is found in a specific step of the approach.
We see that $p_{best}$ is most frequently found on step~1, yet steps~2 and~3 also contribute to finding the best prefix in many cases.
Taken together, steps~2 and~3 contribute 21.7\% and 17.0\% of all $p_{best}$ prefixes for the open-source functions and Stack Overflow datasets, respectively.

Overall, these results indicate that each step in \name{} consistently adds to its effectiveness, with step~1 being the most important.
The low number of prefixes in $P$ indicates that \name{} effectively determines a small number of prefixes that maximize coverage, which is useful to keep the number of executions performed in a any downstream applications of \name{} manageable.

\subsubsection{Trees of Explored Prefixes}
\name{} generates the prefixes that form a tree.
In total, every tree of prefixes could contain up to 210 nodes: 10 nodes from step~1, 100 nodes from step~2, and 100 nodes from step~3.
We investigate how many of these nodes are visited, on average, before \name{} terminates, e.g., because it has already fully covered the given snippet.
Figure~\ref{fig:prefixes_exploration} shows the distribution of the number of prefixes explored in total (``all''), and in each step of the approach.
We also show the number of prefixes in the set $P$.
On average, \name{} explores 70 prefixes for the open-source functions dataset, and only two of these prefixes end up in $P$, i.e., are required to achieve maximum coverage.
The maximum observed size of $P$ is twelve. 
Across the 1,000 snippets in the open-source functions dataset, \name{} achieves full coverage for 558 on step~1. Then, 442 go to step~2, and 392 go to step~3.
Figure~\ref{fig:prefixes_exploration_b} shows the corresponding results on the Stack Overflow dataset.
Here, \name{} explores 39 prefixes, and only one is added to $P$, on average.
The maximum observed size of $P$ is four.
Out of the 462 snippets, \name{} achieves full coverage for 410 on step~1. Then, 52 go to step~2, and 166 go to step~3.

\subsubsection{Impact of Resolving Dependencies}
As described in Section~\ref{sec:install_dependencies}, \name{} uses pipreqs to identify and install missing dependencies.
On the 1,462 snippets used in our evaluation, pipreqs is invoked 1,216 times and succeeds in 1,206 of these invocations. 
The ten failures are caused by syntax errors on the predicted prefixes. 
After identifying dependencies with pipreqs, \name{} tries to install them using pip install.
In total, pip install gets invoked 48 times, as we do not try to install the same dependency twice, which is successful in 38 out of the 48 cases. 
The failures are caused by versions of the libraries suggested by pipreqs that are not available for the version of Python we use, i.e., Python 3.8.

To understand the importance of resolving dependencies, we investigate the effectiveness of \name{} without pipreqs.
For cost reasons, we only consider GPT-4o mini.
The results are in Table~\ref{tab:effectiveness_per_step}, in the lines containing ``w/o pipreqs''.
We observe that for both datasets, not resolving dependencies leads to a decrease in effectiveness, i.e., this part of our approach contributes to achieving high coverage.

\subsection{RQ3: Case Studies}

We qualitatively analyze the strengths and weaknesses of \name{} by inspecting samples of examples.

\subsubsection{Reasons for Higher Coverage}

\begin{figure}
    \centering
    \begin{subfigure}[b]{\linewidth}
        \centering
        \includegraphics[width=0.9\textwidth]{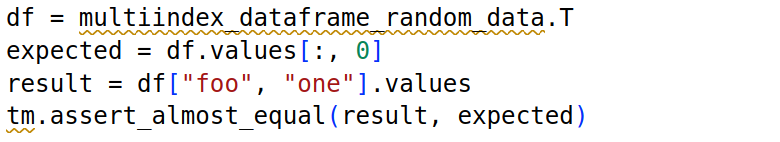}
       % \vspace{-1.7em}
        \caption{Code snippet.}
        \label{fig:case_study_example_snippet}
    \end{subfigure}
  
    \vspace{1em}
    \begin{subfigure}[b]{\linewidth}
        \centering
        \includegraphics[width=0.9\textwidth]{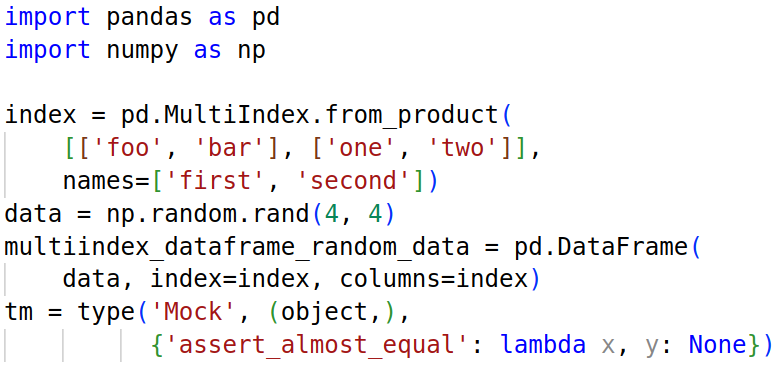}
        % \vspace{-1.7em}
        \caption{Prefix predicted by \name{}.}
        \label{fig:case_study_example_prediction}
    \end{subfigure}
    \caption{Example of adequate imports and usage of dependencies and complex values predicted by \name{}.}
    \label{fig:case_study_example}
\end{figure}

Across all 1,462 analyzed code snippets, \name{} achieves higher coverage 
than LExecutor in 707 code snippets: 561 from the open-source functions dataset and 146 from the Stack Overflow dataset. 
We randomly inspect a sample of 20, 10 from each dataset, leading the the following observations.

\paragraph*{Adequate imports and usage of dependencies} For 15/20 code snippets, \name{} increases coverage by importing a dependency and using it to create adequate values.
LExecutor does not add any imports but always injects a value from a fixed set.

\paragraph*{Complex objects} For 13/20 code snippets, \name{} predicts code that creates complex objects, e.g.,  with \texttt{type("Mock", bases, dict)}.
% following the specification described in in Figure~\ref{fig:response_specification}.
These objects usually contain the attributes and methods used in the code snippet.
Figure~\ref{fig:case_study_example} shows an interesting combination of using imported dependencies and complex objects.
Notice that the code snippet in Figure~\ref{fig:case_study_example_snippet} tries to access a multi-index dataframe with random data.
The prefix predicted by \name{} (Figure~\ref{fig:case_study_example_prediction}) correctly imports \code{pandas} and \code{numpy}, and then instantiates a multi-index dataframe with random values.
Moreover, \name{} assigns to \code{tm} a mock object, which contains an \code{assert\_almost\_equal} method receiving two arguments, as used in the last line of the given snippet.

\paragraph*{Diverse primitive values} For 11/20 code snippets, \name{} predicts domain-specific primitive values,
e.g., meaningful string and integers.
This differs from LExecutor, which predicts values from a fixed set only.

\paragraph*{Multiple paths covered} For 1/20 code snippets, the multi-step algorithm of \name{} yields multiple snippets, that cover different paths.
Since LExecutor makes only one prediction, it fails to fully cover any examples with multiple paths. %and does not reason about errors from its predictions nor it contains a line coverage feedback.

\subsubsection{Reasons for Lower Coverage}

\begin{figure}
    \includegraphics[width=0.2\linewidth]{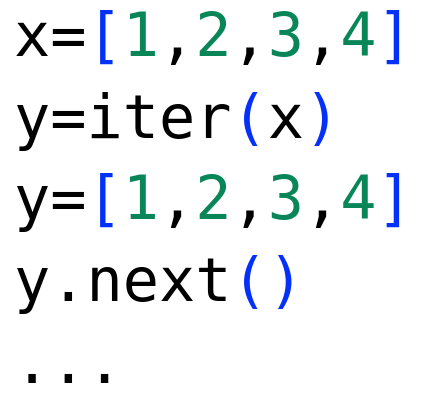}
    % \vspace{-1.7em}
    \caption{Example of problematic code.}
    \label{fig:case_study_problematic_code}
\end{figure}

Across all 1,462 code snippets \name{} has lower coverage 
than LExecutor in 58 code snippets: 47 from the open-source functions dataset and 11 from the Stack Overflow dataset. We randomly selected and inspect 10 snippets from each dataset, with the following observations.

\paragraph*{Missing dependencies and language versions} 12/20 code snippets have problems with missing dependencies.
In these cases, the LLM predicts a correct prefix, but its execution fails because \name{} fails to install the correct dependency.
Moreover, one of the code snippets simply fails because it requires an older version of Python than the one used in our experimental environment.

\paragraph*{Problematic code} 4/20 code snippets are problematic and cannot be fully executed, no matter what the model predicts.
Figure~\ref{fig:case_study_problematic_code} shows an example.
Notice that initially \code{y} is an \code{iterable} and then is 
changed to a \code{list}, which does not contain the \code{next()} method. Therefore, the fourth line will always fail.
Given the the corresponding error message in step~2, \name{} creates a prefix with \code{y = iter([1, 2, 3, 4])}. However, as by design we always add the predictions to the top of the file, \code{y} will always be a list 
when \code{y.next()} is called.

\paragraph*{Unparseable string} We specify the output of the model to be a JSON according to Figure~\ref{fig:response_specification}.
This helps the model to produce predictions that are often more structured and easier to parse than plain text. However, for three of 
the 20 code snippets, the LLM consistently produced strings on the JSON responses, across all samples we get from the model, 
that could not be parsed.

\subsection{RQ4: Diversity of Values}

As observed in RQ3, one of the reasons for the effectiveness of \name{} is the diversity of values predicted by the LLM.
We further study this diversity by analyzing how many unique types and values the approach predicts.
To this end, we consider each of the prefixes predicted by \name{}.
For each prefix, we identify the undefined variables in the corresponding code snippet and the values assigned to them in the prefix.
We consider two primitive values to be equal if they have the same string representation.
For non-primitive values, e.g., complex objects, we consider two values to be equal if they share the same set of attributes and methods, as determined by Python's built-in \code{dir} function.
We then count the number of unique types, and the number of unique values per type, across all prefixes predicted by \name{}.

\begin{table}[t]
    \centering
    \caption{Diversity of values predicted by \name{}.}
    \label{tab:values_diversity}
    \begin{tabular}{lr}
    \hline
    Type            & \multicolumn{1}{l}{Unique values} \\ \hline
    \rowcolor[HTML]{CCD7E6} 
    str             & \textbf{2962}                              \\
    \rowcolor[HTML]{CCD7E6} 
    list            & \textbf{2567}                              \\
    \rowcolor[HTML]{CCD7E6} 
    Mock            & \textbf{1383}                              \\
    \rowcolor[HTML]{CCD7E6} 
    type            & \textbf{1083}                              \\
    \rowcolor[HTML]{CCD7E6} 
    dict            & \textbf{936}                               \\
    \rowcolor[HTML]{CCD7E6} 
    object          & \textbf{919}                               \\
    \rowcolor[HTML]{E2E9F3} 
    set             & 657                               \\
    \rowcolor[HTML]{E2E9F3} 
    MockSelf        & 406                               \\
    \rowcolor[HTML]{E2E9F3} 
    ContextVar      & 342                               \\
    \rowcolor[HTML]{E2E9F3} 
    tuple           & 322                               \\
    \rowcolor[HTML]{E2E9F3} 
    SimpleNamespace & 320                               \\
    \rowcolor[HTML]{E2E9F3} 
    ndarray         & 298                               \\
    \rowcolor[HTML]{E2E9F3} 
    bytes           & 256                               \\
    \rowcolor[HTML]{E2E9F3} 
    Pattern         & 196                               \\
    \rowcolor[HTML]{E2E9F3} 
    module          & 181                               \\
    \rowcolor[HTML]{E2E9F3} 
    Line            & 148                               \\
    \rowcolor[HTML]{E2E9F3} 
    int             & 110                               \\
    \rowcolor[HTML]{EFEFEF} 
    \multicolumn{2}{c}{\cellcolor[HTML]{EFEFEF}...}     \\ \hline
    Total           & 16528                             \\ \hline
    \end{tabular}
\end{table}

\begin{table*}[]
    \caption{Average costs per code snippet imposed by \name{}.}
    \label{tab:expenses}
    \centering
    \resizebox{\textwidth}{!}{
    \begin{tabular}{llrrrrrrrrrrrr}
    \hline
    \multicolumn{1}{c}{Model} & \multicolumn{1}{c}{Dataset} & \multicolumn{4}{c}{Time (seconds)}                                                                                                & \multicolumn{1}{l}{}     & \multicolumn{7}{c}{Price (USD)}                                                                                                                                                                                                            \\ \cline{3-6} \cline{8-14} 
    \multicolumn{1}{c}{}      & \multicolumn{1}{c}{}        & \multicolumn{1}{l}{Step 1}   & \multicolumn{1}{l}{Step 2}  & \multicolumn{1}{l}{Step 3}  & \multicolumn{1}{l}{All}      & \multicolumn{1}{l}{}     & \multicolumn{2}{c}{Step1}                                          & \multicolumn{2}{c}{Step 2}                                      & \multicolumn{2}{c}{Step 3}                                    & \multicolumn{1}{l}{All}       \\ \cline{3-6} \cline{8-14} 
    \multicolumn{1}{c}{}      & \multicolumn{1}{c}{}        & \multicolumn{1}{l}{}         & \multicolumn{1}{l}{}        & \multicolumn{1}{l}{}        & \multicolumn{1}{l}{}         & \multicolumn{1}{l}{}     & \multicolumn{1}{l}{Input}         & \multicolumn{1}{l}{Output}     & \multicolumn{1}{l}{Input}      & \multicolumn{1}{l}{Output}     & \multicolumn{1}{l}{Input}     & \multicolumn{1}{l}{Output}    & \multicolumn{1}{l}{}          \\ \cline{1-2} \cline{8-14} 
    GPT4o                     & Open-source functions       & \cellcolor[HTML]{CCD7E6}13.7 & \cellcolor[HTML]{CCD7E6}2.7 & \cellcolor[HTML]{CCD7E6}2.2 & \cellcolor[HTML]{CCD7E6}\textbf{18.6} & \cellcolor[HTML]{CCD7E6} & \cellcolor[HTML]{CCD7E6}0.002     & \cellcolor[HTML]{CCD7E6}0.019  & \cellcolor[HTML]{CCD7E6}0.056  & \cellcolor[HTML]{CCD7E6}0.106  & \cellcolor[HTML]{CCD7E6}0.132 & \cellcolor[HTML]{CCD7E6}0.109 & \cellcolor[HTML]{CCD7E6}\textbf{0.425} \\
                              & Stack Overflow snippets     & \cellcolor[HTML]{E2E9F3}5.2  & \cellcolor[HTML]{E2E9F3}0.7 & \cellcolor[HTML]{E2E9F3}1.9 & \cellcolor[HTML]{E2E9F3}\textbf{7.8}  & \cellcolor[HTML]{E2E9F3} & \cellcolor[HTML]{E2E9F3}0.001     & \cellcolor[HTML]{E2E9F3}0.006  & \cellcolor[HTML]{E2E9F3}0.012  & \cellcolor[HTML]{E2E9F3}0.013  & \cellcolor[HTML]{E2E9F3}0.119 & \cellcolor[HTML]{E2E9F3}0.059 & \cellcolor[HTML]{E2E9F3}\textbf{0.212} \\ \hline
    \begin{tabular}[c]{@{}l@{}}GPT4o-\\ mini\end{tabular}                & Open-source functions       & \cellcolor[HTML]{CCD7E6}14.2 & \cellcolor[HTML]{CCD7E6}3.8 & \cellcolor[HTML]{CCD7E6}3.2 & \cellcolor[HTML]{CCD7E6}\textbf{21.2} & \cellcolor[HTML]{CCD7E6} & \cellcolor[HTML]{CCD7E6}6.67x10-5 & \cellcolor[HTML]{CCD7E6}0.0007 & \cellcolor[HTML]{CCD7E6}0.0018 & \cellcolor[HTML]{CCD7E6}0.0045 & \cellcolor[HTML]{CCD7E6}0.004 & \cellcolor[HTML]{CCD7E6}0.004 & \cellcolor[HTML]{CCD7E6}\textbf{0.016} \\
                              & Stack Overflow snippets     & \cellcolor[HTML]{E2E9F3}4.4  & \cellcolor[HTML]{E2E9F3}0.4 & \cellcolor[HTML]{E2E9F3}2.2 & \cellcolor[HTML]{E2E9F3}\textbf{7.0}  & \cellcolor[HTML]{E2E9F3} & \cellcolor[HTML]{E2E9F3}3.62x10-5 & \cellcolor[HTML]{E2E9F3}0.0003 & \cellcolor[HTML]{E2E9F3}0.0005 & \cellcolor[HTML]{E2E9F3}0.0008 & \cellcolor[HTML]{E2E9F3}0.004 & \cellcolor[HTML]{E2E9F3}0.003 & \cellcolor[HTML]{E2E9F3}\textbf{0.008} \\ \hline
    \end{tabular}
    }
\end{table*}

In total, for the predictions for the 1,462 snippets on the two datasets, \name{} predicted 1,120 unique types and 16,528 unique values. 
These numbers are in stark contrast to the 23 fixed values predicted by LExecutor.
We attribute this difference to the LLM's ability to predict context-dependent and domain-specific values that fit naturally to the given code.
Table~\ref{tab:values_diversity} presents the most commonly predicted types, ordered by the number of unique values. Interestingly, there is a mix of primitive 
and non-primitive types. 
The type with most predicted unique values is \code{str}, with 2,962 unique values. The prefixes predicted by \name{} in 
Figure~\ref{fig:running_example_snippet} illustrate the importance of domain-specific strings.
Another commonly predicted type is \code{type}, which means that \name{} dynamically creates objects on the fly 
based on the content of the code snippet. %and following our response specification described in Figure~\ref{fig:response_specification}.
An example of a predicted \code{type} is presented in Figure~\ref{fig:case_study_example_prediction}.
Another interesting observation are the 181 unique values for type \code{module}.
This type refers to all imported dependencies, such
as \code{pandas} and \code{numpy} in the prefix in Figure~\ref{fig:case_study_example_prediction}.

\subsection{RQ5: Efficiency and Costs}

Table~\ref{tab:expenses} shows the efficiency and costs of \name{} by measuring (i) the time it takes to execute a code snippet, 
and (ii) the monetary expenses to query the LLMs, based on OpenAI's pricing for GPT4o and GPT4o-mini as of July 2024.
On average, the first step is the most time-consuming. We attribute this to the time required to install dependencies on the prefixes
predicted in step~1. Most times, the prefixes predicted in steps~2 and~3 contain imports previously predicted in step~1, and hence, are already installed.
Overall, on average, \name{} takes 18.6 and 21.2 seconds to execute on an open-source function, and 7.8 and 7 seconds to execute on a
Stack Overflow code snippet with GPT4o and GPT4o-mini, respectively. 
 
As presented on the right side of Table~\ref{tab:expenses}, the price to query both models in \name{} increases along the steps
for both datasets. OpenAI's pricing depends on the amount of tokens consumed and produced by the model.
The amount of tokens in the prompts is higher as the steps in \name{} increase.
On average, \name{} with GPT4o costs USD 0.425 to execute on an open-source function and USD 0.212 to execute on a 
Stack Overflow code snippet.
Using the smaller, less expensive GPT4o-mini models reduces costs significantly, with USD 0.016 to execute on an open-source function and USD 0.008
to execute on a Stack Overflow code snippet.

These results allows for several observations.
First, both \name{}'s execution time and monetary costs depend on the amount of missing values, as the 
open-source functions dataset contains almost double the amount of missing values of the Stack Overflow dataset.
Second, there is a clear trade-off between the cost of a model and its effectiveness.
\name{} with GPT4o is much more expensive than with GPT4o-mini. However, the higher expense is paid with 
higher effectiveness, as presented in Table~\ref{tab:overall_effectiveness}.
For example, on the open-source functions, using the newer model increases costs by 27x, while increasing coverage from 79\% to 84\%.
Third, another trade-off exists between the costs of letting the approach perform all three steps and the benefits obtained in terms of higher coverage.
While the time taken by \name{} increases only slightly in steps~2 and~3, the costs increase significantly.
For example, when using GPT4o on the open-source functions dataset, 94.8\% of the total costs are imposed by steps~2 and~3, which contribute only 14.3\% of the total coverage achieved.
These trade-offs are important to consider when using \name{} in practice, as they provide users with the flexibility to choose the most suitable model and number of steps to balance costs and effectiveness.

\section{Threats to Validity}

Our evaluation is based on two datasets: open-source functions and Stack Overflow snippets.
While these datasets are diverse and representative of real-world scenarios, they may not cover all possible types of Python code.
Additionally, the effectiveness of \name{} might vary with different programming languages.
Following prior work~\cite{lexecutor}, we use line coverage as our primary metric of effectiveness.
While line coverage is a widely accepted measure, it does not capture other important aspects of code execution, such as execution time and memory usage.
Our experiments are conducted using specific versions of the LLMs (GPT-4o and GPT-4o-mini), and future updates to these models could impact the reproducibility of our results.
To mitigate this threat, we make logs of running our experiments available.
Likewise, our results may not generalize to other LLMs.
We reduce this threat by evaluating on the currently best available model (GPT4o), but also on a smaller, less performant model (GPT4o-mini).
Finally, the results depend on the specific prompts used to query the LLMs.
We design these prompts to describe the problem to the LLM in a way similar to how a human would understand it, and we used preliminary, small-scale experiments to tune the prompts.
Other prompts could lead to different results.
To enable future work to build on our results, we provide the prompts used in our replication package.

\section{Related Work}

\paragraph*{Arbitrary code execution}
Prior work explores ways of executing arbitrary pieces of code. Micro-execution~\cite{Godefroid2014} enables executing arbitrary x86 code
by injecting binary values into memory on demand in a virtual machine.
Underconstrained symbolic execution~\cite{Ramos2015} applies symbolic execution to individual functions in isolation.
Forced execution~\cite{Kim2017} forces the execution of uncovered paths.
LExecutor~\cite{lexecutor} enables the execution of arbitrary code snippets using a neural model.
Our work differs from the above by predicting code prefixes that create missing values to maximize coverage.

\paragraph*{Test generation and fuzzing}
Test generation and fuzzing automatically produce tests and inputs to dynamically trigger software errors.
Many approaches use traditional techniques, e.g., symbolic execution~\cite{Cadar2008,Godefroid2005,Sen2005}
or search-based algorithms~\cite{Fraser2011a,pynguin}. Recent work~\cite{Lemieux2023,Pizzorno2024,Deng2023,icse2024-Fuzz4All} 
rely on AI and LLMs.
Both fuzzing and test case generation assume that the target code is complete and executable, e.g. in an environment with all dependencies
installed, and that the input values are provided at well defined locations.
In contrast, \name{} enables the execution of incomplete code and predicts values to be used at arbitrary locations.

\paragraph*{Automated program repair}
APR aims to automatically generate patches for buggy programs.
Recently, researchers have started to apply LLMs for APR~\cite{Xia2023,xia2022,Xia2023a,Bouzenia2024}.
Unlike APR, our approach does not modify the to be executed target code snippet.
Instead, we produce prefixes that are added at the top of the code snippet to enable its execution.

\paragraph*{Learning and LLMs in software engineering}
Learning-based approaches and LLMs have been applied to various software engineering tasks~\cite{NeuralSoftwareAnalysis}, including
type prediction~\cite{Hellendoorn2018,icse2019,fse2020,Allamanis2020},
test generation~\cite{Lemieux2023,schafer2023empirical,Ryan2024,Pizzorno2024},
fuzzing~\cite{Deng2023,icse2024-Fuzz4All},
code completion~\cite{Chen2021,Nie2023,arXiv2024_De-Hallucinator}, and
automated repair~\cite{Chen2021d,Xia2023a,Ye2024,Silva2024,Hossain2024} and coding agents~\cite{Bouzenia2024,Yang2024a,Zhang2024a,Tao2024}.
Our work is the first to propose an LLM-based approach for learning-guided cod execution.

\section{Conclusions}

Motivated by the recurring problem of executing snippets of code, this paper presents \name{}, a novel approach toward learning-guided execution.
Starting from a code snippet, the approach iteratively creates a tree of code prefixes aimed at initializing any undefined references, running the code snippet without errors, and covering as many lines of the code snippet as possible.
Our empirical evaluation shows clear improvements over the previous state of the art, raising line coverage from 59--75\% to 82--84\%. 
We envision \name{} to provide a basis for various dynamic analysis applications, as well as a starting point for developer's trying to understand and debug code.

\section{Data Availability}

Our replication package is available at~\cite{artifacts}.

\bibliographystyle{IEEEtran}
\bibliography{references,referencesMP}

\end{document}